\documentclass{webofc}
\usepackage[varg]{txfonts}   
\begin{document}
\title{Equation of state and strangeness in neutron stars}
%
\subtitle{ - role of hyperon-nuclear three-body forces -}

\author{\firstname{Wolfram} \lastname{Weise}\inst{1}\fnsep\thanks{\email{weise@tum.de}}
}

\institute{Physics Department, Technical University of Munich, 85748 Garching, Germany
          }

\abstract{A brief survey is presented of our present understanding of the equation-of-state of cold, dense matter and the speed of sound in the interior of neutron stars, based on the constraints inferred from observational data. The second part focuses on strangeness in baryonic matter and the role of hyperon-nuclear two- and three-body forces,  with reference to the "hyperon puzzle" in neutron stars and possible scenarios for its solution.
  
}
\maketitle
\section{Introduction}
\label{intro}

The structure of highly compressed baryonic matter matter at zero temperature is still largely unknown.  Neutron stars are the extreme systems of choice to learn about the properties of strongly interacting cold matter at high baryon densities.  Their detailed composition, ranging from nuclear degrees of freedom at low densities to various possible forms of quark matter at high densities, is subject to ongoing speculations. 

Within the last decade,  the observational data base for neutron stars has expanded substantially.  Its detailed analysis sharpened the empirical constraints on the equation-of-state (EoS) of the matter deep inside the core of such objects.  In particular,  the discovery of two-solar-mass neutron stars implied that the EoS,  i.e.  pressure $P(\varepsilon)$ as function of energy density $\varepsilon$,  must be sufficiently stiff in order to support such heavy compact stars against gravitational collapse.  Some previously discussed simple forms of exotic matter (quark matter,  kaon condensation, ...) were thus excluded when their corresponding EoS's turned out to be too soft and unable to satisfy the stability conditions.

Viewing neutron stars as systems primarily composed of neutrons in beta equlibrium with a small fraction of protons,  it has long been considered that the formation of hyperons through weak processes  might become energetically favourable at sufficiently high densities.  However, it was soon realized that when adding these strangeness degrees of freedom,  the EoS would again become too soft and unable to support $2 M_\odot$ neutron stars. This started a discussion under the keyword {\it hyperon puzzle}, which is still continuing and will be a main theme in this presentation.  To frame this discussion in a broader context,  it is useful first to summarize the present state of knowledge about the neutron star EoS as inferred from the observational data.  Of particular interest is the speed of sound in the core of neutron stars.  Its behaviour as a function of energy density is a sensitive indicator for phase transitions or continuous changes of degrees of freedom (crossovers) in the star's composition. 

\section{Equation-of-state of neutron star matter}
\label{sec-2}
\subsection{Observational constraints}
\label{sec-2.1}

{\it Neutron star masses and radii}.  The information about the masses of the heaviest neutron stars derives primarily from precise Shapiro time delay measurements of pulsars orbiting in binary systems with white dwarfs.  Three such massive objects, PSR J1614–2230 \cite{Demorest2010,Fonseca2016,Arzoumanian2018}, PSR J0348+0432 \cite{Antoniadis2013} and PSR J0740+6620 \cite{Cromartie2020,Fonseca2021},  have been established: 
\begin{eqnarray}
	&\text{PSR J1614–2230} \qquad &M = 1.908 \pm 0.016 \, M_\odot ~, ~ \label{eq:ShapiroMass1}\\
	&\text{PSR J0348+0432} \qquad &M = 2.01 \pm 0.04 \, M_\odot ~, \label{eq:ShapiroMass2}\\
	&\text{PSR J0740+6620} \qquad &M = 2.08 \pm 0.07 \, M_\odot ~. \label{eq:ShapiroMass3}
\end{eqnarray}
Radii of neutron stars together with their masses can be inferred from X-ray profiles of rotating hot spot patterns measured with the NICER telescope stationed at the ISS,  also in combination with other multimessenger data.  Two representative neutron stars have been investigated in this way and analysed by two independent groups using different methods.  Results are summarized in Table \ref{tab-1}.
\begin{table}[tp]
\centering
\caption{Masses of two neutron stars together with radii inferrred from NICER measurements at 68\% credible level.}
\label{tab-1}     
\begin{tabular}{llll}
\hline
 & Mass $M \,[M_\odot$]& Radius $R$ [km] & Ref. \\
\hline\\
\vspace*{0.2cm}
PSR J0030+0451 & $1.44^{+0.15}_{-0.14}$ & $13.02^{+1.24}_{-1.06}$ &\cite{Miller2019}\\
 & $1.34^{+0.15}_{-0.16}$ & $12.71^{+1.14}_{-1.19}$ &\cite{Riley2019}
\vspace*{0.2cm}
\\\hline\\
\vspace*{0.2cm}
PSR J0740+6620 & $2.08 \pm 0.07$ & $13.7^{+2.6}_{-1.5}$ & \cite{Miller2021}\\
 & $2.072^{+0.067}_{-0.066}$ & $12.39^{+1.30}_{-0.98}$ & \cite{Riley2021}
\vspace*{0.2cm}
\\\hline
\end{tabular}
\end{table}
These combined mass and radius data are now included as prime input in setting constraints for the neutron star EoS.\\
\\
{\it Binary neutron star mergers and tidal deformabillities}.  Gravitational wave signals produced by the merger of two neutron stars in a binary have been detected by the LIGO and Virgo collaborations.  These signals are interpreted using theoretical waveform models that depend on the mass ratio of the two neutron stars, $M_2/M_1$, and a mass-weighted combination of their tidal deformabilities,  $\Lambda_1$ and $\Lambda_2$:
\begin{equation}
	\bar{\Lambda} = \frac{16}{13}\frac{(M_1 + 12M_2)M_1^4\Lambda_1 + (M_2 + 12M_1)M_2^4\Lambda_2}{(M_1+M_2)^5}~.
\end{equation} 
So far,  two binary neutron star merger events, GW170817 \cite{Abbott2019} and GW190425 \cite{Abbott2020},  were observed,  yielding the following constraints on $\bar{\Lambda}$ at the 90\% level:
\begin{align}
	\text{GW170817} : ~~  \bar{\Lambda}=320^{+420}_{-230}~~, \qquad\qquad
	\text{GW190425} : ~~\bar{\Lambda} \leq 600 ~.
\end{align}
The first one of these events (GW170817) was further evaluated together with electromagnetic signals \cite{Fasano2019}. The following masses and tidal deformabilities of the individual neutron stars in the binary were reported:
\begin{align}
	M_1 = 1.46^{+0.13}_{-0.09}\,M_\odot \qquad & \Lambda_1 =255^{+416}_{-171} ~,\nonumber\\
         M_2 = 1.26^{+0.09}_{-0.12}\,M_\odot \qquad & \Lambda_2 =661^{+858}_{-375} ~.
\end{align}
The tidal deformabilities of the individual neutron stars are evidently still subject to large uncertainties,  but they nonetheless provide further independent data for statistical inference procedures.

\subsection{Inference of the sound velocity and the EoS in neutron stars}
\label{sec-2.2}

\begin{figure}
\centering
\includegraphics[width=10cm,clip]{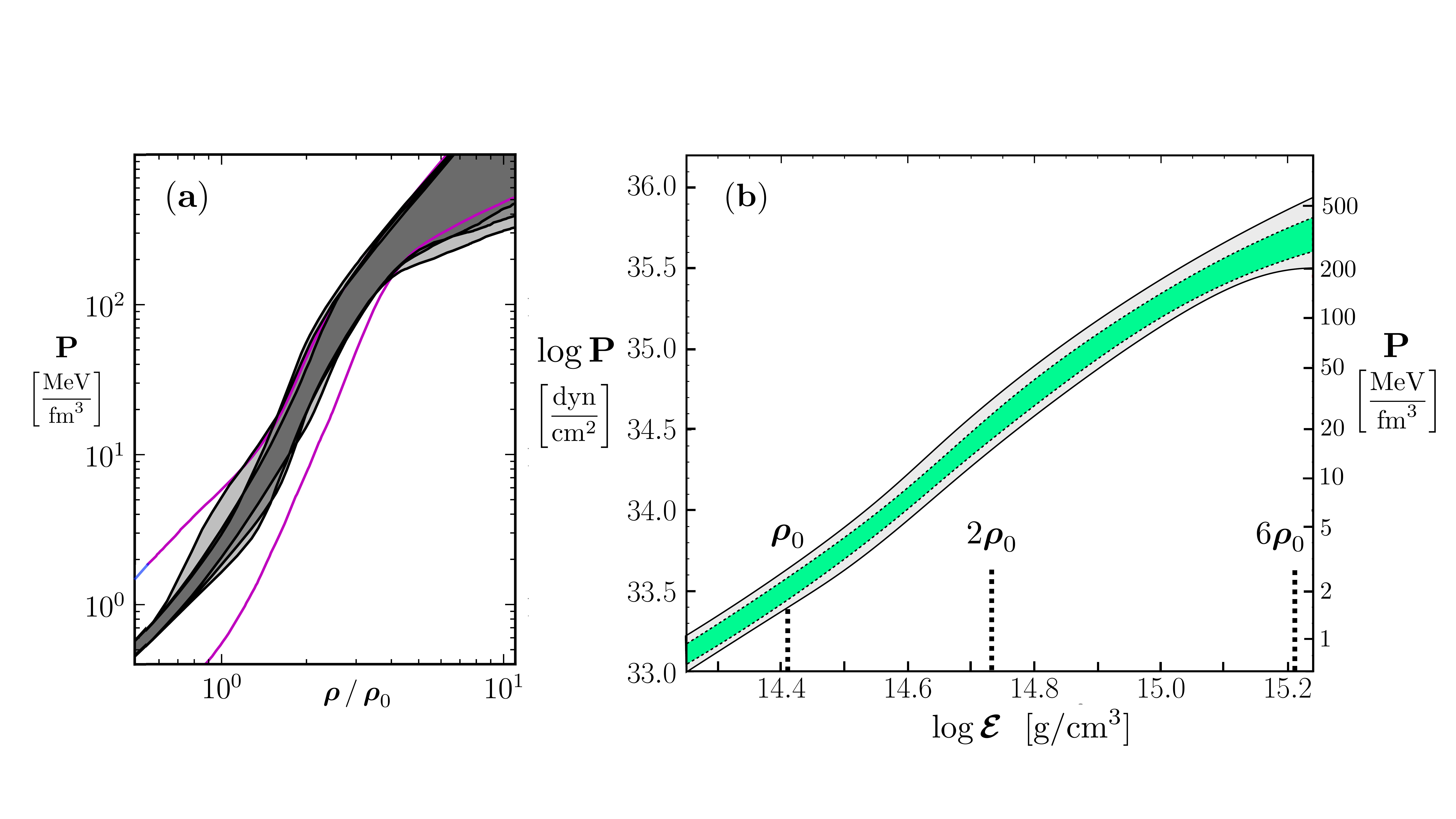}
\caption{Two recent examples of deduced EoS of neutron star matter using Bayes inference methods with data sets described in Section \ref{sec-2.1} as input: (a) pressure $P$ as function of baryon density $\rho$ (in units of $\rho_0 = 0.16$ fm$^{-3}$) \cite{Essick2020} and (b) as function of the energy density $\varepsilon$ \cite{Raaijmakers2021}. }
\label{fig-1}       
\end{figure}

In recent years the data listed in the previous subsection have been incorporated in a variety of studies using Bayesian statistics to infer the equation-of-state of neutron star matter and related properties. Two representative examples of such results are shown in Figure\,\ref{fig-1}.  Constraints at low baryon densities are commonly imposed by implementing results from nuclear chiral effective field theory (ChEFT) in the range $\rho \simeq 1 - 2\,\rho_0$ around the equilibrium density, $\rho_0 = 0.16$ fm$^{-3}$, of normal nuclear matter.  Notably,  the density profiles computed for neutron stars in the mass range $M \sim$ 1.4 - 2.1 $M_\odot$ reach baryon densities that typically do not exceed $\rho \sim 6\,\rho_0$ in the central cores of even the most massive stars,  where central pressures $P \gtrsim 300$ MeV/fm$^3$ are required to ensure stability.

\begin{figure}
\centering
\includegraphics[width=8.5cm,clip]{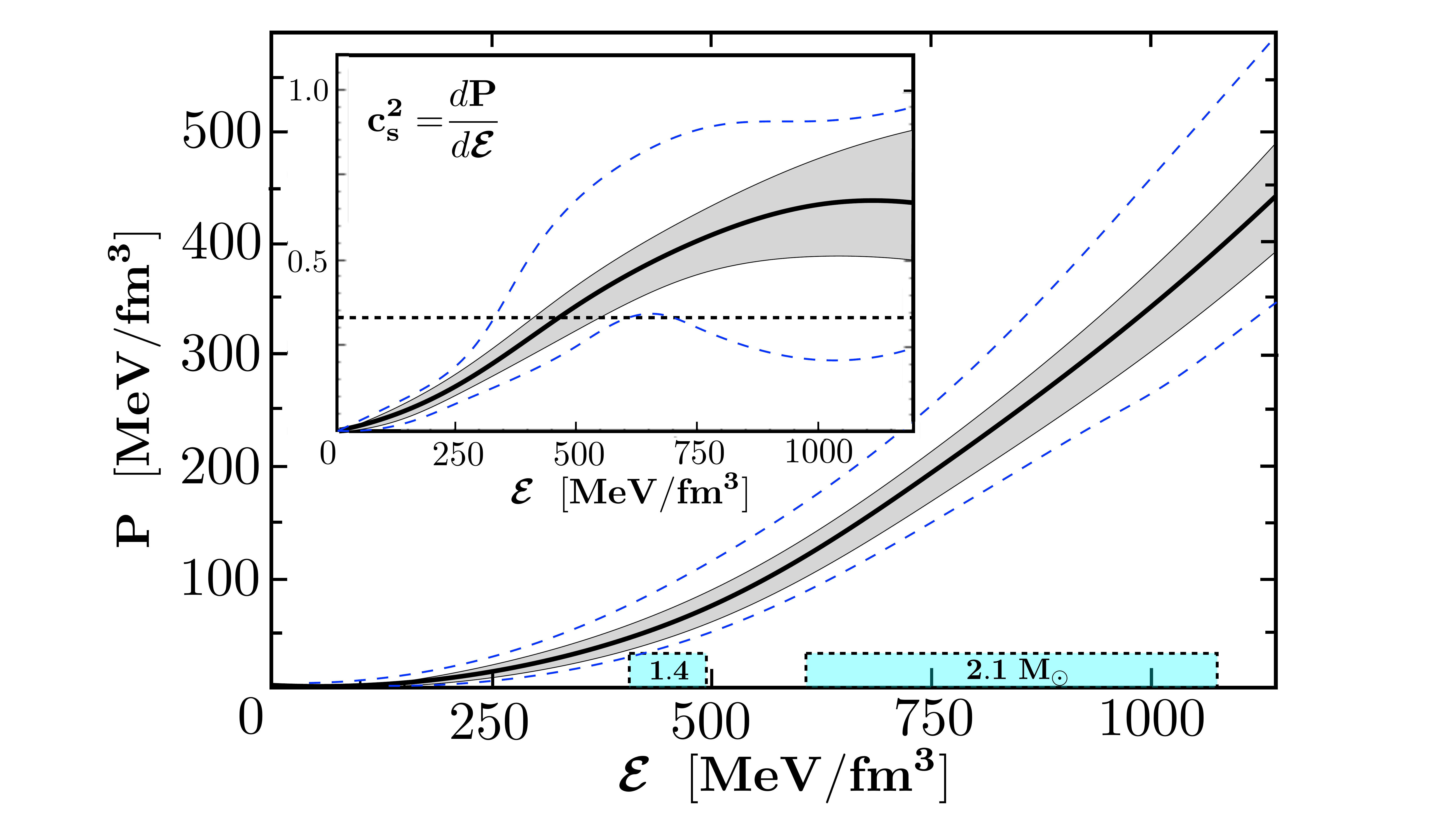}
\caption{Pressure $P(\varepsilon)$ as function of energy density and squared sound speed $c_s^2(\varepsilon)$ of neutron star matter using Bayesian inference with data listed in Section \ref{sec-2.1}. Displayed are medians (solid curves),  68\% (grey) and 95\% (dashed) credible bands. The sectors on the lower axis mark the ranges of central energy densities realised in the cores of 1.4 and 2.1 $M_\odot$ neutron stars,  respectively.  Adapted from \cite{Brandes2022}.}
\label{fig-2}       
\end{figure}

An advanced Bayes inference procedure has recently been performed \cite{Brandes2022} with special focus on the (squared) sound velocity,  $c_s^2(\varepsilon) = {\partial P(\varepsilon)/\partial \varepsilon}$,  from which the EoS is determined as 
\begin{equation}
P(\varepsilon) = \int_0^\varepsilon d\varepsilon' c_s^2(\varepsilon')~.
\end{equation}
Results are shown in Figure \ref{fig-2}. At 68\% credibility level,  it can be concluded that the speed of sound in neutron stars systematically exceeds the canonical value for an ultrarelativistic Fermi gas (the conformal limit),  $c_s^2 = 1/3$,  at energy densities $\varepsilon \gtrsim 0.5$ GeV/fm$^{-3}$ corresponding to baryon densities $\rho \gtrsim 3\,\rho_0$.  This observation is of some significance,  recalling that a first-order phase transition in the core of a neutron star would be signaled by a rapid drop of $c_s$ to zero.  Quantifying likelihood categories in terms of Bayes factors,  one finds that there is strong evidence \cite{Brandes2022} that $c_s^2$ stays at values larger than 0.1 for neutron stars with $M \lesssim 2\,M_\odot$.  This indicates that a first-order phase transition in the core of even the heaviest observed neutron stars is unlikely.  On the other hand,  a continuous crossover at high baryon density is not ruled out.

\begin{figure}[tp]
	\begin{center}
		\includegraphics[height=36.5mm]{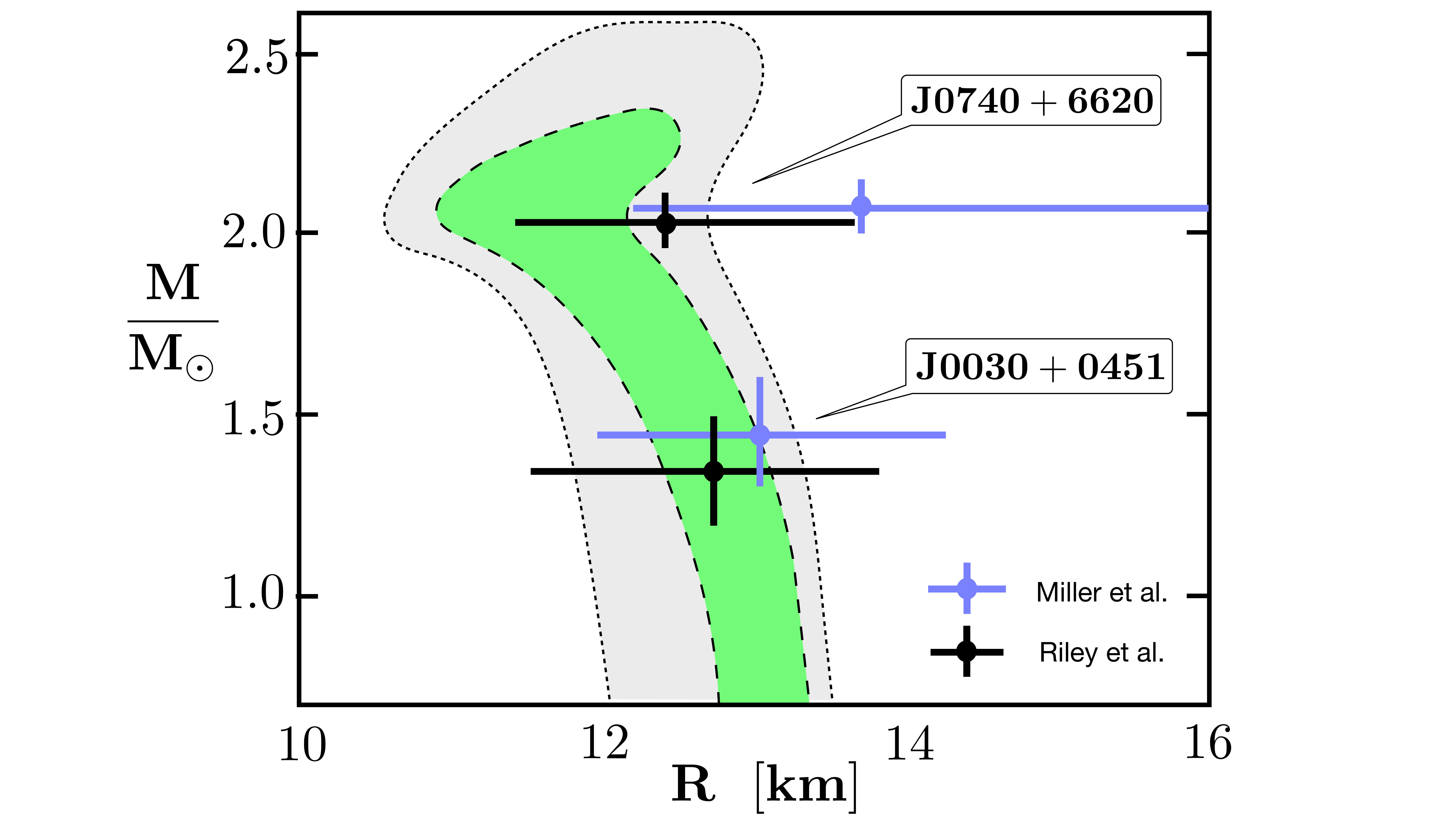} 
		\includegraphics[height=37.5mm]{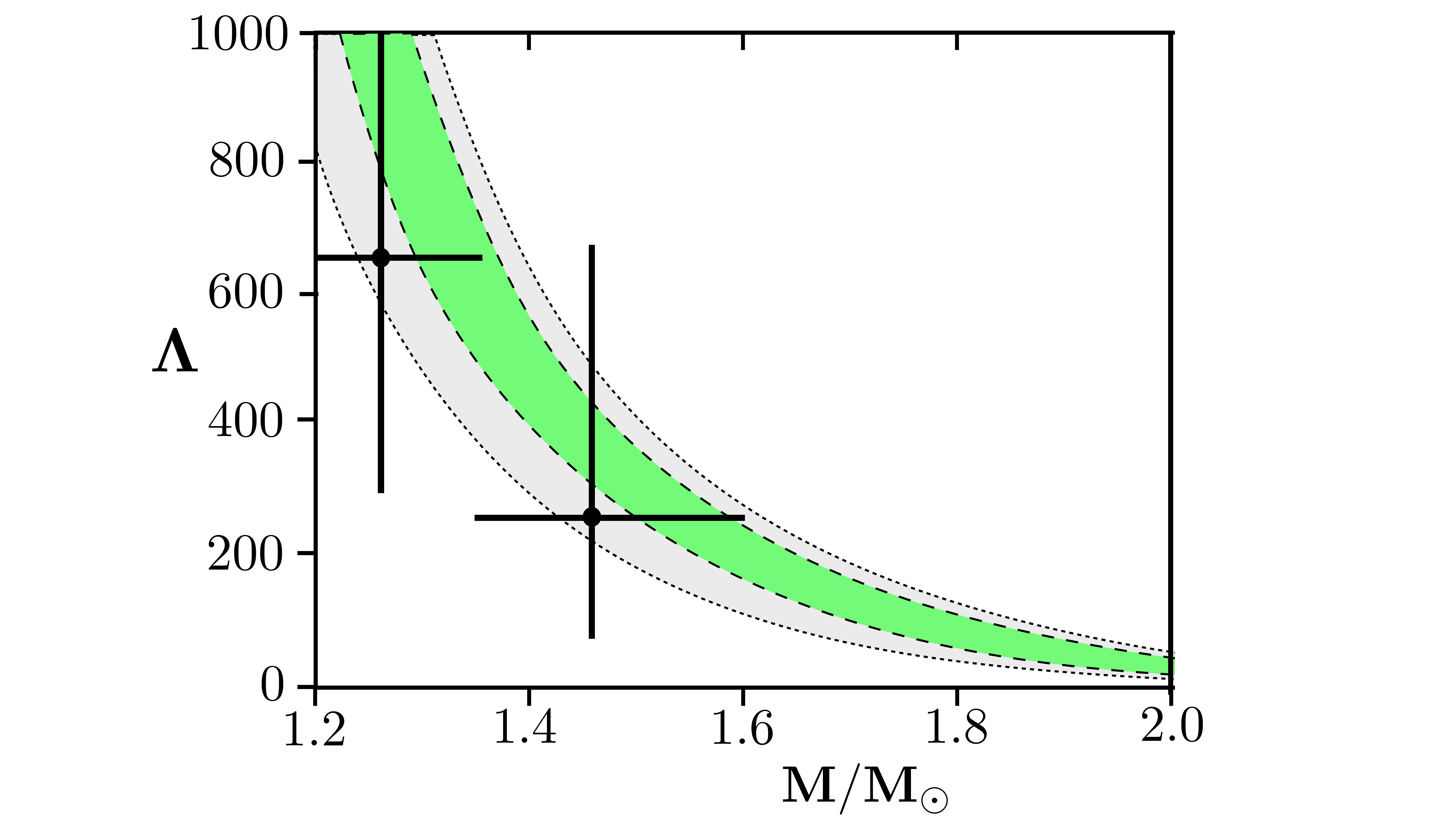} 
\sidecaption
\caption{Inferred 95\% and 68\% credible bands for the mass-radius relation (left) and the mass dependence of the tidal deformabilty $\Lambda$ (right) of neutron stars \cite{Brandes2022} based on the empirical EoS of Fig.\,\ref{fig-2}.  Data for masses and radii from \cite{Miller2019, Miller2021, Riley2019, Riley2021}.  Tidal deformabilty data (GW170817) from \cite{Fasano2019}.}
\label{fig-3}  
\end{center}     
\end{figure}

Figure \ref{fig-3} shows the mass-radius relation and tidal deformabilties for neutron stars as deduced from the inferred EoS and in comparison with data.  The most probable radii are in the range 11-13 km.  A tendency towards slightly smaller inferred radii for the heaviest $2\,M_\odot$ objects as compared to those listed in Table \ref{tab-1} appears to be favoured when the GW merger events are incorporated in the Bayesian analysis.

\subsection{Theoretical frameworks and models}
\label{sec-2.3}

While the advanced statistical methods applied to the expanded observational neutron star data base sharpen the constraints on the equation-of-state of dense matter,  no information is provided in this way about the physical degrees of freedom that govern the microscopic composition and dynamics of the neutron star.  A variety of models exists that can roughly be grouped into three categories:

a) Models based on {\it nuclear degrees of freedom} (nucleons,  pions and short-range correlations) using many-body theory with realistic two- and three-nucleon interactions.  A time-honored example is the APR equation-of-state \cite{APR1998}.  Spontaneously broken chiral symmetry of low-energy QCD is a guiding principle,  realised through chiral EFT calculations \cite{Hebeler2013, Holt2017, Huth2021} applicable up to baryon densities $\rho \lesssim 2\,\rho_0$.  A similar scheme in combination with non-perturbative functional renormalisation group methods can be further extended to higher densities \cite{Drews2017}.  An important point to be aware of when working with such chiral approaches is the following: the frequently used mean-field approximation typically leads to an unphysical first-order chiral phase transition at baryon densities $\rho\lesssim 3\,\rho_0$,  well within the density range realised in neutron stars.  However,  fluctuations beyond mean-field tend to convert this phase transition into a continuous crossover \cite{Brandes2021} that is shifted to densities quite far beyond even those expected in the centers of heavy $2\,M_\odot$ stars.

b) Models operating with {\it nucleons and hyperons},  using advanced many-body methods and including interactions of the baryon SU(3) octet with nucleons as input \cite{Djapo2010,Lonardoni2015}.  These models are commonly confronted with the {\it hyperon puzzle} in neutron star matter unless additional repulsive mechanisms are introduced,  e.g. through hyperon-nuclear many-body forces, in order to maintain the necessary stiffness of the EoS.  An interesting variant in this class of approaches is a chiral quark-meson model \cite{Motta2021} in which the members of the baryon octet are coupled to meson fields with quark bag boundary conditions.  Effective in-medium baryon masses are determined by the non-linear density dependence of a scalar (sigma) field coupled to the baryons.  A key quantity is the scalar polarizability of the baryons that encodes effects of repulsive three-body forces. This model can accomodate $\Lambda$ hyperons in neutron star matter from an onset density $\rho\gtrsim 3\,\rho_0$ upward.  Charged and neutral $\Xi$'s would appear at higher densities,  still maintaining the required stiffness of the EoS in the presence of sufficiently repulsive many-body correlations.  However,  the predicted pattern of the resulting sound speed in the presence of hyperons displays a rapid drop below the conformal bound, $c_s = \sqrt{1/3}$, which raises another issue in view of the empirical constraints discussed previously.  

c) Hybrid models characterised by a {\it hadron-to-quark crossover} connecting nucleonic or baryonic matter at intermediate densities with quark matter at high densities.  Such scenarios have gained significance in recent years under the keywords {\it quark-hadron continuity} or {\it quarkyonic matter} \cite{McLerran2019, Fukushima2020}.  Percolation mechanisms are thought to be at the origin of such a continuous transition.  Quark-antiquark pairs forming meson clouds at lower densities increase their mobility upon compression.  At high densities,  the compact valence quark cores of the baryons begin to overlap.  An example of an EoS constructed within such a picture is given in \cite{Baym2019}.  The quark matter regime at high densities is described by a Nambu - Jona-Lasinio model including pairing interactions.  Strong universal repulsion between vector currents of the quarks is usually introduced to keep the EoS sufficiently stiff.  \\

\begin{figure}
\centering
\includegraphics[width=8.0cm,clip]{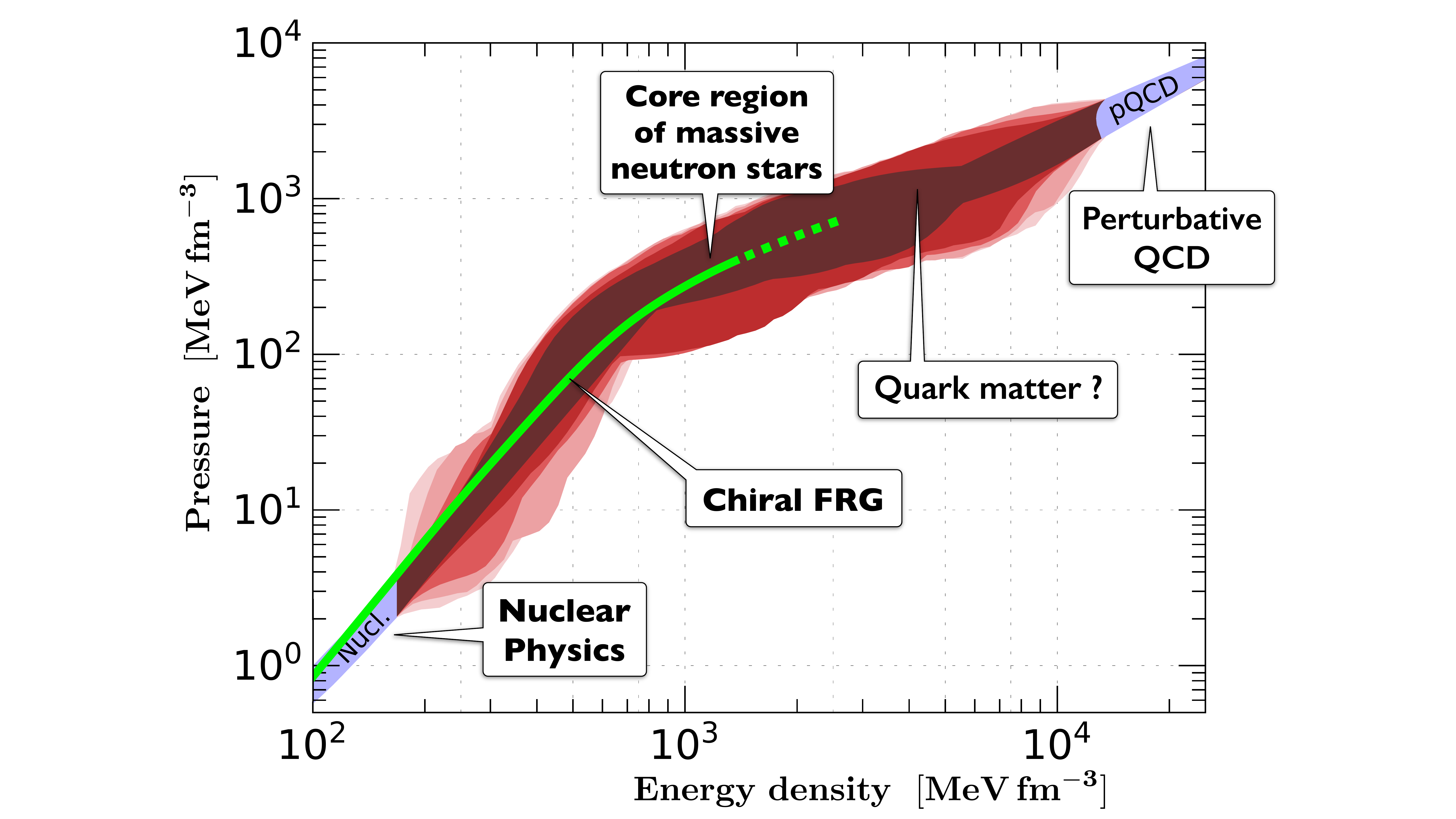}
\caption{Equation-of-state $P({\varepsilon})$ covering neutron star matter and beyond, with low-density constraints from nuclear physics \cite{Hebeler2013, Holt2017, Huth2021},  asymptotic constraints from perturbative QCD and uncertainty bands interpolating between these extremes (adapted from \cite{Annala2020}). Also shown is an EoS curve derived from chiral nucleon-meson field theory combined with functional renormalization group methods (Chiral FRG) \cite{Drews2017}. }
\label{fig-4}       
\end{figure}

Figure \ref{fig-4} may be taken as a pictorial summary of this brief survey,  displaying the EoS with built-in constraints from nuclear physics at low densities,  the perturbative QCD limit at asymptotically high energy densities,  and interpolating between these extremes.  An apparent change of slope in $P(\varepsilon)$ around typical core densities of heavy neutron stars is sometimes interpreted as a possible signature of a hadron-to-quarks phase transition.  But its statistical weight,  judging from the previous inference analyses,  is so far limited.  In fact several other inference results for $P(\varepsilon)$ do not show a pronounced kink structure. 

\begin{figure}
\centering
\includegraphics[width=7.7cm,clip]{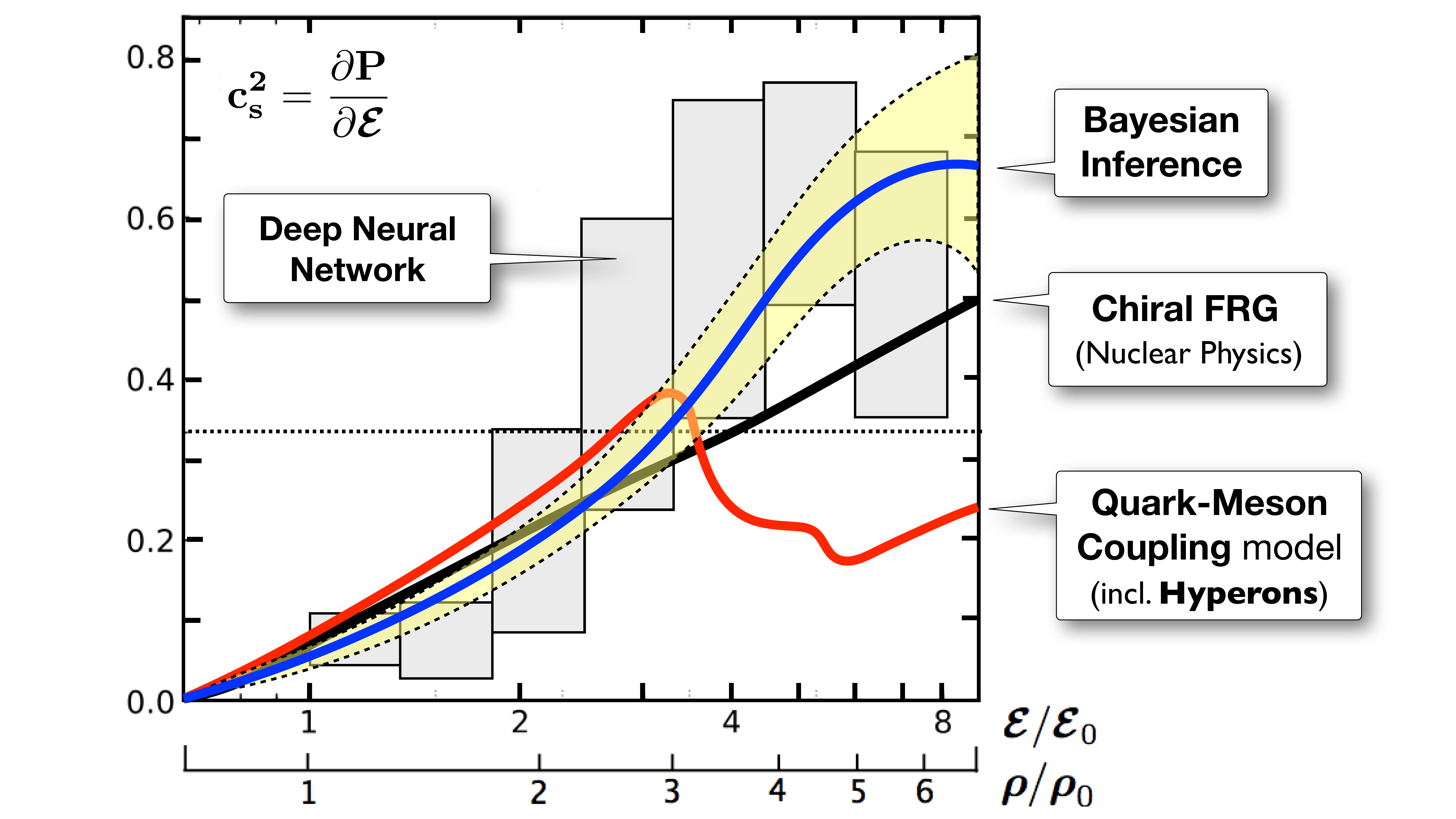}
\caption{Squared speed of sound in neutron star matter as function of energy density (in units of $\varepsilon_0 = 150$ MeV/fm$^3$) and of baryon density (in units of $\rho_0 = 0.16$/fm$^3$). Bayes inference result (68\% credible band) \cite{Brandes2022} is shown side-by-side with data-based mapping of $c_s^2$ using a deep learning network approach \cite{Fujimoto2020}.  For comparison: chiral nucleon-meson field theory combined with functional renormalization group (Chiral FRG) \cite{Drews2017},  and quark-meson coupling model calculation including hyperons \cite{Motta2021}. }
\label{fig-5}       
\end{figure}

As emphasized previously,  the sound velocity in neutron stars is a key quantity in order to further reduce the freedom of choice in models describing the intrinsic structure and composition of dense matter in neutron stars.  This is apparent in Figure \ref{fig-5}.  Shown there is the Bayesian inference posterior of $c_s^2$ from \cite{Brandes2022} together with a reconstruction of the sound speed from neutron star data using a deep learning network scheme \cite{Fujimoto2020}.  These results demonstrate once again the rise of $c_s^2$ beyond the conformal boundary of 1/3 within the range of neutron star densities.  A chiral FRG calculation \cite{Drews2017} based on nucleon and meson degrees of freedom follows this trend (though perhaps with a slightly too soft EoS at the highest densities).  The quark-meson coupling model \cite{Motta2021} including hyperons indicates a pattern that falls out of this scheme.  Obviously there is a strong quest for a further reduction of uncertainties in the sound speed along with future extensions of the neutron star data base.

\section{Strangeness and baryonic matter}
\label{sec-3}
\subsection{Hyperon-nucleon interactions from SU(3) chiral effective field theory}
\label{sec-3.1}

Guided by the symmetry breaking pattern of three-flavor QCD,  chiral SU(3) effective field theory (EFT) is a systematic low-energy expansion, with controlled uncertainties, of interactions involving the baryon and pseudoscalar meson octets.  Unresolved short-distance dynamics is encoded in contact terms and derivatives thereof.  Our present (still limited) quantitative understanding of hyperon-nucleon interactions derives primarily from this framework at next-to-leading order (NLO) \cite{Haidenbauer2013, Haidenbauer2020, Petschauer2020}.  Steps beyond NLO are possible in principle but not yet feasible in practice because of the large number of low-energy constants that need to be determined at higher orders. This requires a much broader empirical data base compared to the one presently available.  Nonetheless the updated (NLO19) version of the interaction  \cite{Haidenbauer2020} has been checked successfully against an increasing amount of data.  It now serves as a prototype force for $\Lambda N$,  $\Sigma N$ and $\Xi N$ two-body processes,  including important coupled-channel dynamics such as $\Lambda N \leftrightarrow \Sigma N$ and $\Xi N\leftrightarrow \Lambda\Lambda$.

{\it Hyperon-nucleon scattering and correlation functions}.  Hyperon-nucleon cross sections measured decades ago at BNL and CERN served as a first step in constraining the theory.  In the meantime interesting new experimental developments provide further tests.  Examples are the $\Sigma^- p \rightarrow \Sigma^- p$ elastic scattering and $\Sigma^- p \rightarrow \Lambda n$ charge exchange scattering measurements at J-PARC (E40) \cite{Miwa2021},  and the femtoscopy studies of hyperon-nucleon correlation functions from $pp$ collisions with ALICE at LHC \cite{Fabbietti2021}.  In particular,  the accurately determined $\Lambda p$ correlation function \cite{Acharya2021} is well reproduced by the NLO19 interaction,  especially at the lowest center-of-mass momenta ($\lesssim 100$ MeV/c) which are not covered by scattering experiments. 

{\it Constraints from hypernuclei}.  Hypernuclear physics has provided a systematic set of binding energies for ground states and excited states of a $\Lambda$ in nuclei from A = 5 to 208.  A phenomenological $\Lambda$-nuclear single-particle potential works well to reproduce the data with a central potential depth $U_\Lambda(r=0) \simeq -30$ MeV \cite{Gal2016}.  Hypernuclear many-body calculations,  using e.g. Brueckner-Hartree-Fock methods \cite{Haidenbauer2020b, Gerstung2020},  have henceforth addressed the question to what extent such a shell-model potential can be understood in terms of underlying $\Lambda N$ two-body forces.  The results of those computations point to a trend, namely that a realistic two-body interaction produces too much binding in the lower shell-model levels of heavy hypernuclei such as $_\Lambda Pb$ \cite{Haidenbauer2020b}.  Similarly,  the $\Lambda$ potential in nuclear matter calculated with the NLO19 two-body force alone is more attractive by almost 20\% than the empirical potential and has its minimum located at a density 25\% higher than $\rho_0 \simeq 0.16$ fm$^{-3}$ \cite{Gerstung2020}. 

This opens the question about the role of $\Lambda NN$ three-body forces. 
The issue is further underlined by a recent phenomenological study \cite{Friedman2022} using an ansatz that combines contributions of $\Lambda$-nucleon two- and three-body forces to the $\Lambda$-nuclear potential: $U_\Lambda(\rho) = U_0^{(2)} [\rho(r)/\rho_0 ]+ U_0^{(3)} [\rho(r)/\rho_0]^2$,  in an optimal fit to the systematics of hypernuclear binding energies of the lowest ($s$ and $p$ shell) orbits.  The best-fit result features a total potential depth of $U_\Lambda(\rho_0) = -(26.5\pm 1.6)$ MeV,  separated into a $\Lambda N$ two-body contribution,  $U_0^{(2)} = -(40.4\pm 0.6)$ MeV,  and a repulsive $\Lambda NN$ three-body piece,  $U_0^{(3)} = (13.9\pm 1.4)$ MeV.  Systematic uncertainties may be underestimated in this work,  but the presence of a substantial repulsive hypernuclear three-body force is nonetheless suggestive.

{\it Hyperon-nuclear three-body forces}. Within the chiral EFT hierarchy,  genuine three-baryon interactions first appear at next-to-next-to-leading order (NNLO) \cite{Petschauer2016}. A complete calculation would involve a large number of parameters, too large for a reliable determination given the limited amount of data.  However,  an approximation known to work well in the three-nucleon sector,  namely $\Delta(1232)$ dominance in intermediate states of the three-body mechanism,  can easily be generalized to decuplet dominance in SU(3) chiral EFT,  now with $\Sigma^*$ and $\Delta$ intermediate virtual excitations.  In this way the three-body terms are promoted from NNLO to NLO in the chiral counting,  in line with the NLO treatment of the two-body forces.  This leaves three coupling parameters of the three-body interaction terms.  Assuming SU(3) symmetry,  one of these coupling constants is already fixed by the $\Delta \rightarrow \pi N$ decay width.  The remaining two parameters can be constrained by comparison with the empirical $\Lambda$-nuclear single-particle potential \cite{Gerstung2020}.

\subsection{Hyperons in dense baryonic matter and neutron stars}
\label{sec-3.2}

Equiped with a $\Lambda NN$ three-body force from chiral SU(3) EFT,  one can now proceed to investigate its role in nuclear and neutron star matter.  

\begin{figure}
\centering
\includegraphics[width=9.0cm,clip]{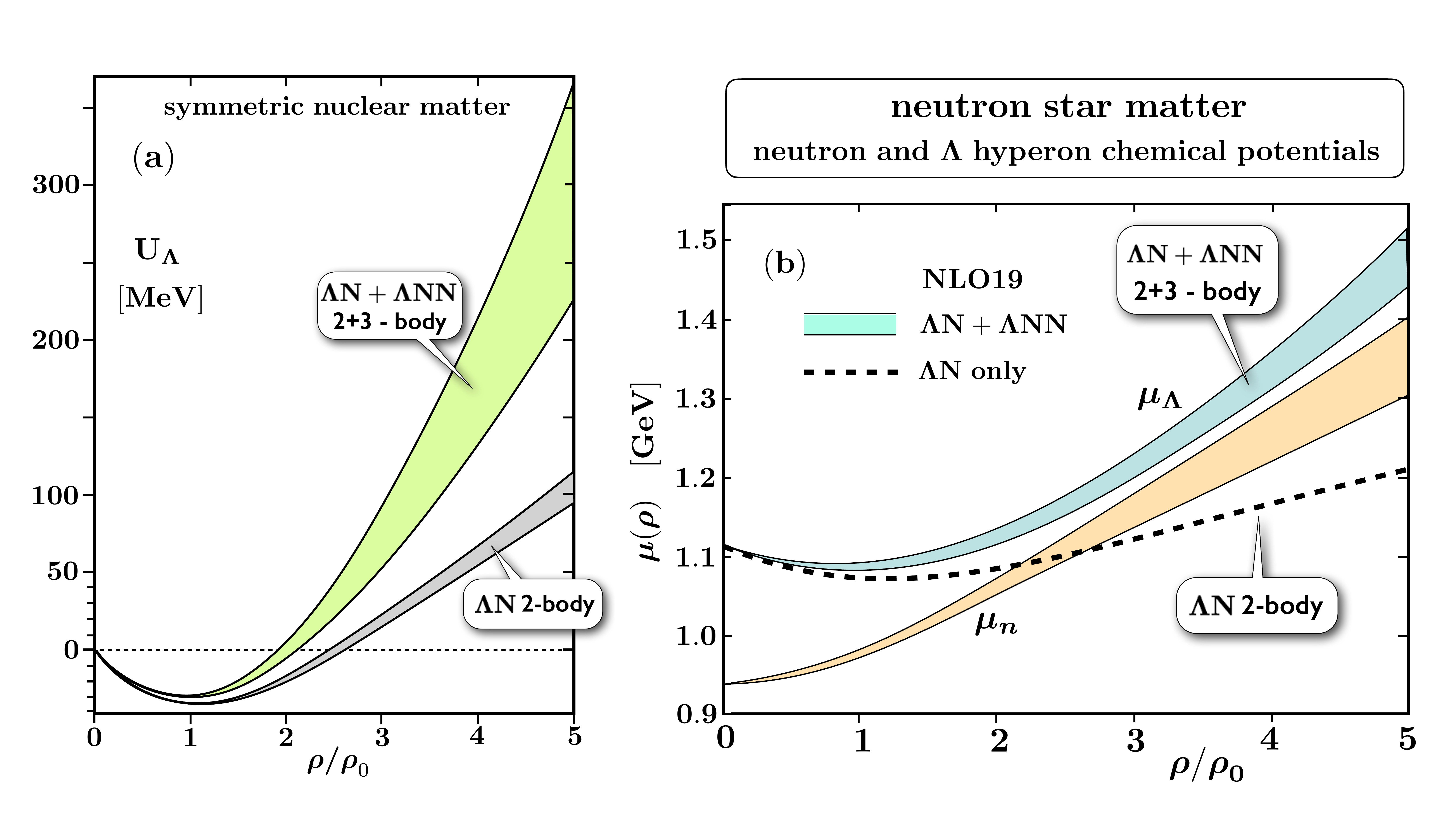}
\caption{(a) Single-particle potential $U_\Lambda$ of a $\Lambda$ hyperon in symmetric nuclear matter as a function of baryon density in units of $\rho_0 = 0.16$ fm$^{-3}$.  Results of Brueckner-Hartree-Fock calculations using NLO19 two-body $\Lambda N$ interaction, in comparison with additional $\Lambda NN$ three-body forces as input.  Three-body parameters have been fixed to the empirical $U_\Lambda(\rho_0) \simeq -30$ MeV.  (b) Chemical potentials of $\Lambda$ and neutron, $\mu_\Lambda$ and $\mu_n$ in beta-equlibrated neutron star matter calculated with the same input as in (a).  The neutron chemical potential is computed using the chiral nucleon-meson model combined with functional renormalization group methods \cite{Drews2017}.  In both Figures the bands give an impression of estimated uncertainties.  Adapted from \cite{Gerstung2020}.}
\label{fig-6}

\end{figure}

{\it Density dependence of the $\Lambda$-nuclear potential}.  With increasing baryon density beyond $\rho\sim\rho_0$ the repulsive nature of the $\Lambda NN$ three-body interaction is expected to become progressively more prominent relative to the contribution of the $\Lambda N$ two-body interaction,  because it enters the energy density with a higher power of density.  This is shown in Figure\,\ref{fig-6}(a) for the $\Lambda$-nuclear single-particle potential,  $U_\Lambda(\rho)$, in symmetric nuclear matter.  A smilar behaviour is observed for a $\Lambda$ hyperon in neutron matter.  At high densities,  $\rho \gtrsim 3\,\rho_0$,  the three-body contribution to $U_\Lambda$ begins to be dominant over the two-body part with its linear density dependence characteristic of a Hartree potential generated by the short-distance $\Lambda N$ interaction.

{\it Hyperon puzzle in neutron stars}.  Given the energy density $\varepsilon(\rho_i)$ of neutron star matter as a function of the densities $\rho_i$ of baryonic constituents (neutrons, protons and possibly $\Lambda$ hyperons),  the individual chemical potentials are determined by $\mu_i = \partial\varepsilon / \partial\rho_i$.  The onset for the appearance of hyperons in a neutron star through weak interaction processes converting neutrons into $\Lambda$'s is marked by the condition that the $\Lambda$ chemical potential equals that of the neutrons: $\mu_\Lambda = \mu_n$.

The density dependence of the potential $U_\Lambda$ is reflected in $\mu_\Lambda$.  Hence the repulsive $\Lambda NN$ three-body interaction shows up prominently in the $\Lambda$ chemical potential at high baryon densities.  Figur\,\ref{fig-6}(b) displays results of a Brueckner-Hartree-Fock calculation of $\mu_\Lambda$ for neutron star matter in comparison to $\mu_n$ \cite{Gerstung2020}.  If only two-body $\Lambda N$ interactions (NLO19) are included,  hyperons start appearing already at densities $\rho \simeq 2-3\,\rho_0$. The EoS becomes too soft and unable to support two-solar-mass neutron stars.  However,  repulsive $\Lambda NN$ three-body forces raise $\mu_\Lambda$ to a level such that it does not match the neutron chemical potential any more: the appearance of $\Lambda$'s in the core of the neutron star is prohibited.  At the same time the neutron sector (with inclusion of many-body correlations) features sufficiently strong repulsion by itself to satisfy the stabilty conditions for heavy neutron stars.

Of course,  uncertainties still exist concerning the detailed quantitative behaviour of the three-body terms in the energy density,  related to approximations involved e.g.  in the simplified octet dominance ansatz.  Hence the hyperon puzzle is not yet solved,  but the results so far obtained point into a qualitatively promising direction.

\section{Conclusions and outlook}
\label{sec-4}

Significant advances have been achieved in the physics of pulsars and the analysis of observational data,  resulting in improved constraints on the equation-of-state and the sound speed at high densities in neutron star interiors.  Progress has also been made concerning the role of strangeness in neutron star matter,  with emphasis on hypernuclear three-body forces and their possible role in addressing the {\it hyperon puzzle}.  The repulsion provided by the $\Lambda NN$ three-body interaction at high baryon densities is potentially capable of blocking the appearance of $\Lambda$ hyperons in the density range of even the heaviest observed ($\sim 2\,M_{\odot}$) neutron stars.  

In this context,  as well as in the general quest for identifying the active degrees of freedom in dense baryonic matter,  an important issue is the further detailed exploration of the sound velocity in the center of neutron stars.  Presently available constraints, though still with relatively large uncertainties,  indicate that the sound speed systematically exceeds the conformal bound,  $c_s = \sqrt{1/3}$,  in the stellar core.  This sets stringent limits on a possible softening of the equation-of-state through the appearance of new phases or additional degrees of freedom in the core.

Further insights into the role of hyperons in neutron star matter will require yet a more quantitative evaluation of the subtle balance between two- and three-body forces involving hyperons.  Much improved and expanded sets of high-statistics two-body hyperon-nucleon scattering and reaction data,  together with correlation functions from precise femtoscopy measuremets,  will provide further quantitative constraints on the theory (e.g.  enabling extensions of the SU(3) chiral EFT approach beyond NLO).

This is in turn necessary to define an accurate baseline for a more detailed investigation of three-body effects and beyond in hypernuclear few- and many-body systems.  Future developments in high-resolution hypernuclear spectroscopy,  combined with advanced methods in many-body theory,  will be mandatory for further progress in specifying the input at normal nuclear densities and at the same time reducing uncertainties in the extrapolation to higher densities.  A foreseen mass resolution of better than 0.5 MeV in a next generation of hypernuclear spectroscopy experiments with $(\pi^+, K^+)$ reactions at J-PARC's projected HIHR beamline \cite{Nakamura2021},  and with $(e,e' K^+)$ measurements at JLab and MAMI,  will be important steps towards these goals.  \\

{\it Acknowledgements:} Special thanks to Len Brandes, Dominik Gerstung and Norbert Kaiser for collaborations and discussions on the topics of this report.  Work supported in part by the DFG Cluster of Excellence ORIGINS.

\end{document}